\documentclass[aps,article,nofootinbib,groupedaddress]{revtex4}
\usepackage[dvipdfmx]{graphicx}
\usepackage{graphicx}
\usepackage{amsmath,amssymb}
\usepackage{hyperref}
\usepackage{appendix}
\usepackage{bm}
\usepackage{color}

\hypersetup{
    colorlinks=true,
    linkcolor=red,
    citecolor=blue,
}

\def\be{\begin{equation}}
\def\ee{\end{equation}}
\def\ba{\begin{eqnarray}}
\def\ea{\end{eqnarray}}

\newcommand{\clvv}{C_{l}^{VV}}

\newcommand{\lcdm}{\Lambda \rm CDM}
\newcommand{\pop}{\rm Pop\ III}

\newcommand{\nbhalo}{\rho_b^{\rm halo}}
\newcommand{\rvir}{R_{\rm vir}}
\newcommand{\ob}{\Omega_{b}}
\newcommand{\omc}{\Omega_{c}}
\newcommand{\tage}{t_{\rm age}}
\begin{document}

\title{Circular Polarization of the CMB: A probe of the First stars}

\author{Soma De$^{1}$, Hiroyuki Tashiro$^{2}$}

\affiliation{$^1$School of Earth and Space Exploration, Arizona State University, Tempe, AZ 85287, USA  \\
$^2$Physics Department, Arizona State University, Tempe, AZ 85287, USA  
}

\begin{abstract}
While it is revealed that 
the Cosmic Microwave Background~(CMB) is linearly polarized at 10~\% level, it is
predicted that there exists no significant intrinsic source for circular
polarization~(CP) in the standard cosmology.
However, during the propagation through a magnetised plasma,
the CP of the CMB could be produced via the Faraday conversion~(FC).
The FC converts a pre-existing linear
polarization into CP in presence of a 
magnetic field with relativistic electrons.
 In this paper, we focus on the FC due to supernova remnants of the first stars, 
also called Pop III stars.
 We derive an
analytic form for the angular power spectrum of the CP of the CMB
generated by the general FC.
We apply this result to the case of the FC
triggered by explosions of the first stars and evaluate the angular
power spectrum, $C_l^{VV}$.
%
We show that the amplitude of $l(l+1)C_l^{VV}/(2\pi)$ $ >10^{-2}\mu$K$^2$ for
$l>100$, with only one Pop III star per halo, the age of Pop III SN 
remnants as $10^4$ years and
frequency of CMB observation as $1$~GHz. We expect the CP of the CMB to be
a very promising probe of the yet unobserved first stars, primarily due
to the expected high signal along with  
an unique frequency dependence.
\end{abstract}
\maketitle

\section{Introduction}

Observations of the Cosmic Microwave Background (CMB) are essential in
modern cosmology.
In particular, 
a precise measurement of the CMB polarization is one of the major goals 
for ongoing and future CMB observations.
Theoretical studies of the CMB polarization predict a 10\% level in linear
polarization under standard
cosmology~\cite{Kosowsky:1994cy,Zaldarriaga:1996xe,Kamionkowski:1996ks,Hu:1997hv}.
The linear polarization of the CMB can be produced 
by anisotropic Thomson scattering around the epoch of recombination~\cite{1968ApJ...153L...1R,1983MNRAS.202.1169K}.
Since the first detection of CMB polarization anisotropy by DASI~\cite{Kovac:2002fg},
several observations have measured the angular power spectrum of the
polarization and the cross-correlation with the CMB temperature
anisotropies~(e.g.,~Ref.~\cite{Hinshaw:2012aka} for one of recent works).
These observational results are consistent with the theoretical
predictions of the cosmological observables that follow from the standard $\Lambda$CDM model.
On the other hand, the circular polarization (CP, hereafter) of the CMB is usually assumed to be zero,
because there is no generation mechanism at the epoch of recombination
within the standard cosmology.

However, the CP of the CMB can be created in the
free-streaming regime $after$ the epoch of recombination.
One of such generation mechanism is the Faraday conversion~(FC) which is
formalized by the generalized Faraday rotation.
Due to the FC,
the linear polarization of the CMB can be converted to the CP with the presence of relativistic magnetized
plasma~\cite{sazonov1969,1977ApJ...214..522J}.
The FC could be expected when the CMB propagate through
relativistic magnetized plasma in galaxy clusters.
Ref.~\cite{Cooray:2002nm} has shown that 
the FC due to galaxy clusters might be able to create
the CP at the level of $10^-9$
at frequencies of 10 GHz.


The CP of the CMB can be generated by other
mechanism. Mohammadi have investigated the generation of the CP of the CMB through their scattering with the cosmic
neutrino background~\cite{Mohammadi:2013dea}. 
Giovannini has shown that the curvature perturbations can produce the CP 
with the presence of primordial magnetic fields around the last
scattering surface~\cite{Giovannini:2009ru, Giovannini:2010yy}.
Sawyer has discussed the CP due to photon-photon
interactions mediated by neutral hydrogen background~\cite{Sawyer:2012gn}.
In addition, some new physics effects can induce the CP of
the CMB~\cite{Alexander:2008fp,Bavarsad:2009hm, Motie:2011az}.

Recently Mainini et al. have performed the first attempt to detect
the CP of the CMB since~'90~\cite{Mainini:2013mja}. They have improved the upper limit
on the degree of the CP, which is between $5 \times
10^{-4}$ and $0.4 \times 10^{-4}$ at large angular scales (between $8^\circ$ and $24^\circ$). 
However, this limit is very far from $10^{-9}$ degree of the CP predicted in a cosmological context.

In this paper, we evaluate the CP of the CMB via
the FC in supernova (SN) remnants of the first stars.
The formation of the first stars is a important milestone in the
evolution of the structure formation.
After photon decoupling, overdensity regions began to grow
and collapsed to dark matter halos.
Inside of dark matter halos, formation of luminous objects like stars was not solely driven by
gravity and requires a sufficient amount of 
baryon gas cooling inside of a dark matter halo, to eventually form stars.
These first born stars are thought
to be very massive and are termed as the first stars or alternatively
Pop III stars.
It is believed that Pop III stars formed in small halos
($10^6$-$10^8~{\rm M}_\odot$)
at $z\sim20$-30~(see Refs.~\cite{Bromm:2009uk, Whalen:2012nk} for
recent reviews).
Although Pop III stars are key to early structure formations as the first
luminous objects and the
sources of cosmic reionization and cosmic metal pollution, no Pop
III stars have been directly observed and
there is some debate on their properties including mass range of Pop III
stars.
In the isolation scenario, Pop III star mass is predicted to be
massive, $100$--$500~{\rm M}_\odot$~\cite{Bromm:1999du, Abel:2000tu, Abel:2001pr, Bromm:2001bi,Nakamura:2000ez,Yoshida:2008gn}.

It is known that Pop III stars with mass $>10~{\rm M}_\odot$ cause SNe
at their death.
The detection of SNe of Pop III stars are expected as one of the 
possible probes of first stars as these SNe could be much brighter than
their progenitors or host galaxies. In particular, Pop III stars with
140-260~${\rm M}_\odot$ could 
explode as pair-instability SNe, which is up to
100 times more energetic than Type Ia and Type II SNe~\cite{Heger:2001cd}.
Many works have been done to investigate the observability of Pop III SNe~\cite{Scannapieco:2005zq,Mesinger:2005du,Kasen:2011eh,Pan:2011aa,Hummel:2011qs,Tanaka:2012hn,Whalen:2012sf,Whalen:2012yk,deSouza:2013fna}.
According to these works, SNe of Pop III stars could be found by the James Webb Space
Telescope (JWST) or the
Wide-Field Infrared Survey Telescope (WFIRST).
Additionally, SN remnants of Pop III stars may be also detectable.  
Meiksin and Whalen have found that
SN remnants of Pop III stars in $10^7~{\rm M}_\odot$ halos can produce 
observable radio signatures~\cite{Meiksin:2012wj}.
Oh et al. have shown that
SN remnants of Pop III stars may induce additional CMB temperature
anisotropy through the Sunyaev-Zel'dovich effect~\cite{Oh:2003sa}.

In this paper, we adopt a simple analytic model for the evolution of SN
remnants to study the FC 
in a SN remnant of a Pop III star.
Our aim is to evaluate the anisotropy of the CMB CP. We
calculate the power spectrum of the FC by using the halo formalism~\cite{Seljak:2000gq}, then we compute the angular
power spectrum of the CMB CP. Throughout this paper, we adopt a flat $\Lambda$CDM cosmology,
with $h = 0.7$, $\Omega_c$ = 0.23, and $\Omega_b =0.046$.

\section{Circular polarization due to Faraday conversion}
\label{basics}

Due to the Thomson scattering with the presence of quadrapole temperature anisotropy,
the CMB becomes linearly polarized during the epoch of
recombination.
As the CMB propagates through inter-galactic medium and galaxy clusters, there
are secondary effects which are imprinted on the CMB temperature
anisotropy and
polarization properties. 

One of such secondary effects is the creation of the circular polarization~(CP)
in the CMB through the mechanism of Faraday Conversion~(FC). 
The FC can be understood in the following way.
Consider a linearly polarized electro-magnetic (EM) wave
in a homogeneous magnetized plasma. Let the direction of propagation of
the EM wave be orthogonal to the direction of the external magnetic
field in the plasma. 
This linearly polarized EM wave can be decomposed into two
linear polarized waves with the same phase,
perpendicular and parallel to the external 
magnetic field.
Circular polarization of an  EM wave can be visualized as two linear
polarized waves with a phase difference between the linearly polarized
components.
Charged particles in the plasma are free to move along the magnetic 
field lines and can respond easily to the electric field of the EM wave.
However, the motions of the charged particles perpendicular to the
magnetic field lines are affected by the
magnetic field and their response to the electric field of the EM wave
is now modified. Therefore, a difference arises in the
particle motions between the two orthogonal polarization directions of the
EM field, due to the existence of the magnetic field in the plasma.
This difference translates into a phase difference between two linearly
polarized components of the EM wave parallel and perpendicular to the
external magnetic field. As a result, a circularly polarized wave is
generated.
%

In this section, after giving a brief review of the Stokes parameters,
we formulate the generation of the CP of the CMB due to the FC
to obtain the analytic form of the CP angular power spectrum.

\subsection{Stokes parameters}

First of all, we consider a monochromatic EM wave propagating along
$\hat{z}$. This EM wave is characterized by two mutually perpendicular electric
field components on the $x$-$y$ plane. At a given point in space, the amplitude of the electric field 
vectors pointing along  $\hat{x}$ and $\hat{y}$ respectively are
described by~\cite{kosowsky-easy}
\begin{eqnarray}
E_x&=&E^0_x(t)\cos(\omega t-\phi_x(t)), \nonumber \\
E_y&=&E^0_y(t)\cos(\omega t-\phi_y(t)) .
\end{eqnarray}
The extent of polarization is generally quantified in terms of
the so-called Stokes parameters $I$, $Q$, $U$ and $V$. 
These Stokes parameters for a monochromatic EM wave are
defined as
\begin{eqnarray}
I&=&(E^0_x)^2+(E^0_y)^2, \nonumber \\
Q&=&(E^0_x)^2-(E^0_y)^2, \nonumber \\
U&=&2E^0_xE^0_y\cos(\phi_x-\phi_y), \nonumber \\
V&=&2E^0_xE^0_y\sin(\phi_x-\phi_y) . 
\end{eqnarray}

However, in practice, we measure EM waves at a frequency $\omega $ with
a bandwidth $\Delta \omega$. That is, measured EM waves can be expressed
as  a superposition of many waves around $\omega$. For such EM waves,
the Stokes parameters are obtained by the time averaging of the electric
field components of the EM waves,
\begin{eqnarray}
I&=&\langle (E^0_x)^2 \rangle+\langle(E^0_y)^2\rangle, \nonumber \\
Q&=&\langle (E^0_x)^2\rangle -\langle(E^0_y)^2\rangle, \nonumber \\
U&=& \langle 2E^0_xE^0_y\cos(\phi_x-\phi_y)\rangle, \nonumber \\
V&=&\langle 2E^0_xE^0_y\sin(\phi_x-\phi_y) \rangle  ,
\end{eqnarray}
where the bracket $\langle ~ \rangle$ denotes the time averaging with the
time interval over which the measurement is performed.

As shown in above equations,
the $I$ parameter represents the intensity of the EM waves,
the $Q$ and $U$ parameters are associated to the LP of the
EM waves and the $V$ parameter quantifies the extent of the CP.
The parameters $I$ and $V$ are coordinate-independent~(scalar) and dimensionless true observables.
The parameters $Q$ and $U$ transform under a rotation of the coordinate
system  while
$Q^2+U^2$ is an invariant under the rotation of the axes.
The sign of the parameter $V$ related to the rotating direction of electric
field components on the $x$-$y$ plane.
EM waves with $V>0$ and $V<0$ are called, respectively, right-handed and
left-handed circular polarized waves.
A linearly polarized wave
is a combination of one left circularly and one right circularly
polarized waves with equal amplitudes, that is, $V=0$. A circularly polarized wave is
created when the left and right circularly polarized components have
unequal amplitudes. 

%
%

\subsection{Angular power spectrum of circular polarization}

For the analysis of the CMB anisotropy,
it is useful to perform an angular decomposition of the anisotropic
values in multipole space.
Let $V(\hat {n})$ be the CP at a
given direction $\hat {n}$ on the sky.  Since the Stokes parameter $V$ is a
scalar quantity, it can be expanded in the basis of
scalar spherical harmonics $Y_{lm}(\hat {n})$ in the following way,
\begin{equation}
V(\hat {n})=\sum_{lm}V_{lm}Y_{lm}(\hat {n}).\label{eq:dec_V}
\end{equation}
Using the coefficients $V_{lm}$,
we can write the angular power spectrum of the CMB circular
polarization, $\clvv$, as
\begin{equation}
\clvv=\frac{1}{2l+1}\sum_{m}V_{lm}V^*_{lm}.\label{eq:def_clvv}
\end{equation}

Let us consider the CP of the CMB due to the FC.
The observed CP is given in terms of the Stokes $V$
parameter by~\cite{sazonov1969}
\begin{equation}
\label{eq:eq_V}
V(\hat{n})=-2\int_{r_*}^{0}dr ~ U(r, \vec{x},\hat{n})
\alpha(r,\vec{x},\hat{n}, \hat{b}),
\end{equation}
where $r$ is the comoving distance, $*$ denotes the value at the last
scattering surface and $U(\vec{x},\hat{ n})$ is the
Stokes parameter at a comoving space position $\vec{x}$ and observation direction
$\hat n$ with $\vec{x} = r \hat n$.
Here $\alpha(r,\vec{x},\hat{n}, \hat{b}) $ is the FC rate with the
magnetic field direction $\hat {b}$, which we will discuss
in more detail in the next section.
As shown in Eq.~(\ref{eq:fcrate}) of the appendix, $\alpha(r,\vec{x},\hat{n}, \hat{b}) $ can be
decomposed as
\begin{equation}
\alpha(r,\vec{x},\hat{n}, \hat{b})=
 2\pi  \sqrt{\frac{32 \pi}{15}} 
\int
\frac{d^3k}{(2\pi)^3} {\tilde \alpha}(z,k)
\left(
 {}_{2}Y_{2}^0( \hat n)+ {}_{-2}Y_{2}^0(\hat n)
\right)
\sum_l
(-i)^lj_l(kr)\sum_{m=-l}^{m=l}Y^*_{lm}(\hat{k})Y_{lm}(\hat{n}),\label{eq:a_dec}
\end{equation} 
where
$_{\pm 2} Y_l^m (\hat{n})$ are spin-2 spherical harmonics, $j_l(x)$ is
the spherical Bessel function, 
and, for simplicity, we assume that $\hat {b}$ is $(0,0)$ in a polar
coordinate system for the sky, $\hat{b} =(\theta, \phi)$.

Generally, the stokes parameters for a linear polarization, $U$ and $Q$, can be
expanded as~\cite{Hu:1997hv}
\begin{equation}
\left ( Q \pm iU \right)(r, \vec{x},\hat{n}) =\int \frac{d^3k}{(2\pi)^3} \sum_l
\sum_{m=-2}^{m=2} \left(E_{lm}\pm iB_{lm} \right) _{\pm 2}G_l^m
(\vec{x}, -\hat n),
\label{eq:eb_decomp}
\end{equation}
where $_{\pm 2}G_l^m$ is a mode function for a spin-2 field,
\begin{equation}
\label{eq:glm}
 _{\pm
2}G_{l}^{m} (\vec{x}, \hat{n})=(-i)^{l}\sqrt{\frac{4\pi}{2l+1}} \ _{\pm 2} Y_l^m(\hat{n})e^{i\vec{k}.\vec{x}}.
\end{equation}
In Eq.~(\ref{eq:eb_decomp}), $E_{lm}$ and $B_{lm}$ are
coefficients for so-called E- and B-mode polarizations~\cite{Kosowsky:1994cy,Zaldarriaga:1996xe,Kamionkowski:1996ks,Hu:1997hv}.
For simplicity, we assume that B-mode polarization vanishes hereafter.
In this assumption, the parameter $U$ is given in terms of $E_{lm}$
as 
\begin{equation}
2iU(r, \vec{x},\hat{n}) =\int \frac{d^3k}{(2\pi)^3} \sum_l
\sum_{m=-2}^{m=2} E_{lm}\left( _{2}G_l^m -_{-2}G_{l}^m \right).
\label{eq:u_dec}
\end{equation}

With Eqs.~(\ref{eq:a_dec}) and(\ref{eq:u_dec}), we can decompose the Stokes parameter $V(\hat{n})$ in Eq.~(\ref{eq:eq_V}) in
spherical harmonics as shown in Eq.~(\ref{eq:dec_V}).
After a lengthy calculation, we finally obtain the angular power spectrum
of $V$ as
\begin{equation}
\label{eq:clvv}
\clvv \approx 
\frac{128 }{15 \pi}
\sum_{l' l'' l'''} \sum_{m'=-2}^{m'=2} \sum_{m''+m'''=m}
\int_{r_*} ^0 dr 
\int 
\frac{k^2dk}{(2l'+1) r^2}
P_\alpha
\left(r, \frac{l''}{r}\right)
P_{E_{lm}}
\left(r, k\right)
j_{l'''}^2(kr) 
I_{lm}^2,
\end{equation}
where $P_{E_{lm}}(r, k)$ is the power spectrum of the E-mode polarization, $E_{lm}$, at
a comoving distance $r$, $P_\alpha(r, k)$ is the power spectrum
of the FC rate at $r$,
and $I_{lm}$ is given by Eq.~(\ref{eq:complicated}).
Note that the detailed derivation of Eq.(\ref{eq:clvv}) is in the appendix and $P_\alpha$ has a dimension of (length). 
Eq.~(\ref{eq:clvv}) conveys that,
depending on the power spectrum, $P_\alpha (r,k)$, the E-mode
polarization is converted to the CP.

It is worth discussing the implications carried by
Eq.~(\ref{eq:eq_V}). This equation quantifies a transfer of polarization
from an existing Stokes U into V, via the FC mechanism. 
As we will see in the following
section, $\alpha$ in Eq.~(\ref{eq:eq_V}) depends only on the $magnitude$
of the component of the $total$ magnetic field that is perpendicular to the line
of sight and is situated on the plane of the sky. In the derivation of
Eq.~(\ref{eq:eq_V}), the $\hat{y}$ axis is chosen to be parallel to the
component of the magnetic field on the plane of the sky. Stokes U is
defined to be situated on the x-y plane and makes an angle of $\pi/4$
with respect to the $\hat{y}$. The specification of the coordinate
system is important because Stokes Q and U are coordinate dependent
quantities. Stokes V and I are however, invariants.
This implies if the 
orientation of the magnetic field on the plane of the sky
changes (without a change in the line of sight component of the magnetic
field), then the observer's measure of Stokes Q and U changes in the
context of Eq.~(\ref{eq:eq_V}). This
does not change the measure of V. Please see more on this discussion after we have
introduced the parameter $\alpha$ in the next section.

Another important factor in this context is FR of the EM
wave. FR induces transfer between the Stokes Q and U components.
This happens to due to the rotation of the incoming polarization of the
EM field due to the line of sight magnetic field component.
As the plasma becomes more relativistic, the FR effects in the plasma
decreases ~\citep{beckert}. In this paper, we
have considered a relativistic plasma and hence considered the FR
effects on the EM wave to be insignificant. A more realistic calculation
must involve a full address of the problem involving both the FR and FC
mechanisms.


\section{Faraday conversion from Pop III stars}

Within the $\lcdm$ framework of structure
formations, dark matter halos begin to be formed around redshifts of $z
\sim 20$-$30$.
Inside of the dark matter halos, Pop III stars were born.
Although the final mass of the Pop III stars is determined by several
dynamical feed back processes related the pre-stellar gas, typically it is
estimated to span between 60-300${\rm M}_{\odot}$. 

In this paper, we investigate the CP signals generated when CMB
photons pass through the SN remnants of the $\pop$ stars.
A SN generated due to a $\pop$ star
explosion produces a large outburst of energy~\cite{Whalen:2012sf}
and a shock wave. As the shock wave propagates through the ambient
medium of the explosion, a strong magnetic field and
a large number of relativistic electrons are produced. 
Consequently, CMB photons passing through SN remnants of $\pop$ stars could
be significantly affected by the FC.

Adopting a simple analytic model of the explosion
of a $\pop$ star, first, we estimate the FC induced by one SN remnant of $\pop$
stars. 
In order to estimate the angular power spectra of the CP, $\clvv$, obtained
in the previous section,  we need to calculate the FC power spectrum, $P_{\alpha}$. Based on the halo model, we
evaluate $P_{\alpha}$ due to SN remnants of $\pop$ stars.

\subsection{Faraday Conversion due to a $\pop$ star explosion}

Faraday conversion rate $\alpha$ in Eq.~(\ref{eq:eq_V}) is given by~\cite{sazonov1969}
\begin{eqnarray}
\label{eq:alpha}
\alpha(z,\vec{x},\hat{n}, \hat{b}) &=&\alpha_0\sin(\theta_{B})^{\frac{\gamma+2}{2}},
\nonumber \\
\alpha_0 &=& 
 C_{\gamma} \frac{e^2}{m_ec}
 n_{\rm rel}\epsilon_{\rm
min}(B_{\rm mag})^{\frac{\gamma+2}{2}}
\nu^{-\frac{\gamma+4}{2}},
\end{eqnarray}
where
$\theta_B$ is the angle between the direction of the line of
sight~$\hat{n}$ and the magnetic field direction~$\hat{b}$, $n_{\rm
rel}$ is the number density of relativistic electrons and $\gamma$
denotes the power-law distribution of the relativistic electrons which
described in
terms of the Lorentz factor $\epsilon$ as $n_{\rm rel} (\epsilon) = n_0
\epsilon^{-\gamma}$ between $\epsilon_{\rm min} <\epsilon <
\epsilon_{\rm max}$.
The parameter $C_\gamma $ in Eq.~(\ref{eq:alpha}) is provided by~\cite{sazonov1969},  
\begin{eqnarray}
\label{eq:fc-params}
C_{\gamma}&=&
\begin{cases}
 -2\frac{\gamma-1}{\gamma-2}(\frac{e}{2\pi m_ec})^{\frac{\gamma+2}{2}}
 \bigl[({\nu(z)/\nu_{\rm L}})^{\frac{\gamma-2}{2}}-1\bigr],
 & \gamma \neq 2 \\
    (\frac{e}{2\pi m_ec})^{2} \log ({\nu(z)/\nu_{\rm L}}),
 & \gamma = 2,
\end{cases}
\end{eqnarray}
where $\nu_{\rm L}$ is the characteristic frequency of the synchrotron
emission by electrons at the lower bound $\epsilon_{\rm min}$, which is
represented as
$\nu_{\rm L}=eB/(2\pi\epsilon_{\rm min}m_{\rm e})$. We picked $\epsilon_{\rm min}=100$.
%

When a SN occurs, large amount of energy is injected into the surrounding
gas. As a result, the shock is created and, then, relativistic electrons and
magnetic fields are generated inside the shocked gas (SN remnants).
In order to estimate the FC rate in SN remnants,
we evaluate the number density of relativistic electrons $n_{\rm rel}$ and magnetic
fields $B_{\rm mag}$, assuming the regime at which
the shock front expands adiabatically. This regime is known as the blast wave regime and the
dynamics of this regime is expressed in the Sedov similar solution which
describes a point blast spherical explosion in an ambient
medium~\cite{Sedov1959}.
In this solution, the radius of the shock $r_s$ is given by 
\begin{equation}
\label{eq:rs}
r_s
\sim
2~{\rm pc}
\left(\frac{E_{\rm SN}}{10^{53} \rm erg}\right)^{1/5}
\left(\frac{\Omega_b h^2}{0.0245} \right)^{-1/5}
\left( \frac{1+z}{20} \right)^{-3/5}
\left(\frac{t_{\rm age}}{10^6 ~\rm yr}\right)^{2/5}
,
\end{equation}
where $E_{\rm SN}$ is the energy of the SN explosion
and $\tage$ is the time since the explosion.
In Eq.~(\ref{eq:rs}), we assume that
the shock expands into an ambient medium with the mean baryon mass
density $\bar \rho_b = \bar \rho_{b0} (1+z)^3$ where $\rho_{b0}$ is the mean
baryon mass density at the present, although we will discuss the baryon
density in more detail below.

%

The energy $E_{\rm SN}$ generally depends on a mass of the exploded
$\pop$ star.  Since $\pop$ stars are predicted to be massive,
60-300~${\rm M}_{\odot}$, we assume that
$\pop$ stars have 100~${\rm M}_{\odot}$ mass and explode as pair-instability
SNe which are 100 times more powerful than typical type II SNe.
Therefore, we adopt $E_{\rm SN} = 10^{53}$~ergs in this paper.

The shock wave expands in the interior of a dark matter halo at first.
Then the shock wave spreads out to the outside of the halo, because
$r_s$ could become larger than the size of the halo as $\tage$ increases.
We simply assume that the shock continues to expand
following Eq.~(\ref{eq:rs}) in both the inside and the outside of the
halo. Therefore, we set $\rho_b$ to
\begin{equation}
\label{eq:nb}
    \rho_b= 
\begin{cases}
    \rho_{b}^{\rm halo},& r_s \leq R_{\rm vir}\\
    \bar{ \rho}_b, & r_s > R_{\rm vir},
\end{cases}
\end{equation}
where $R_{\rm vir}$ is the virial radius of the halo and $\bar{\rho}_b$
is the mean baryon
mass density in the universe. 
To obtain the baryon mass density inside of a dark matter halo $\rho^{\rm halo}_b$,
we assume that the baryon mass distributes homogeneously inside the
virial radius.
Accordingly, the baryon mass density in the halo is given by
\begin{equation}
\label{eq:nbhalo}
\nbhalo=\frac{\ob}{\ob+\omc}\frac{3M}{4\pi\rvir^3}.
\end{equation}

The blast wave phase continues until the cooling of the SN remnant
becomes effective. One of the important cooling mechanisms is the inverse
Compton~(IC) scattering. The cooling time, $t_{\rm IC}$, for the IC scattering
is independent of temperature and density of the gas,
\begin{equation}
\label{eq:tic}
t_{\rm IC} = \frac{3 m_e c}{4 \sigma_T \rho_{\rm CMB}}
\approx 1.4 \times 10^7~{\rm yr}~ \left( \frac{1+z}{20} \right)^{-4},
\end{equation}
where $\sigma_T$ is the Thomson cross-section of an electron and
$\rho_{\rm CMB}$ is the CMB energy density. Eq.~(\ref{eq:tic}) sets an
upper limit on $t_{\rm age} \lesssim 10^7~{\rm yr}$, used for the
estimation of $r_s$ in Eq.~(\ref{eq:rs}). 

Let us evaluate the number density of relativistic electrons. Suppose that
the fraction $f_{\rm rel}$ of the explosion energy $E_{\rm SN}$ gets
converted into relativistic energy of electrons.
For a hydrodynamic shock, the inner radius of the shocked regime~(SN remnant) is estimated to be~\cite{Osterbrock}
\begin{equation}
\label{eq:rp}
r_p=r_s\left(\frac{\eta}{\eta-1} \right)^{-1/3}
\end{equation}
This equation is based on the conservation of mass in the shock
enclosed regime, where the density rises by a factor of $\eta$.
For the case of monoatomic gas, which we adopt here, $\eta=4$.
Therefore, the width of a remnant is $r_s - r_p \sim 0.1 r_s$.
We simply assume that the relativistic electrons are confined in the
region between the radius $r_p$ and $r_s$, whose volume is $V_{\rm rem}
= {4\pi (r_s^3-r_p^3)}/3$.
Assuming the power-law distribution of the relativistic electrons as
mentioned above,
we can obtain the normalization $n_0$ of the distribution from
\begin{equation}
 f_{\rm rel} E_{\rm SN} = V_{\rm rem} \int_{\epsilon_{\rm min}}^{\epsilon_{\rm max}}
n_0 m_e c^2 \epsilon^{1-\gamma}.
\end{equation}
In our calculation, we use $\gamma = 2$, $\epsilon_{\rm min}=100$ and $\epsilon_{\rm
max}=300$.
   
For magnetic fields, we also introduce the parameter $f_{\rm mag}$
which denotes the fraction of the energy of a $\pop$ star explosion
into magnetic field energy. The magnetic
field amplitude $B_{\rm mag}$ is obtained by solving the equation
\begin{equation}
\label{eq:bmag}
\frac{B^2_{\rm mag}}{8\pi} V_{\rm rem} =f_{\rm mag}E_{\rm SN}.
\end{equation}
Therefore, the magnetic field amplitude in a Pop III remnant can be
approximately given by the following. 
\begin{equation}
B_{\rm mag} \sim 70~{\rm mG} \left(
 \frac{t_{\rm age}}{10^6 {\rm yr}} \right)^{-3/5} \left( \frac{E_{\rm SN} }{10^{53} {\rm
ergs}} \right)^{1/5} \left( \frac{1+z}{20} \right)^{{9}/{10}} \left( \frac{f_{\rm mag}}{0.1}\right)^{1/2}. 
\end{equation}
Although the parameters $f_{\rm rel}$ and $f_{\rm mag}$ are theoretically
uncertain, we have set $f_{\rm rel} = 0.1$ and $f_{\rm mag}=0.1$ in this
paper.　In these parameters, 
the typical value of $\alpha_0$ is given by the following. 
\begin{equation}
 \alpha_0 \sim 20~{\rm pc}^{-1}~\left(
\frac{t_{\rm age}}{10^6 {\rm yr}} \right)^{-12/5} \left( \frac{E_{\rm SN} }{10^{53} {\rm
ergs}} \right)^{4/5}
\left( \frac{1+z}{20} \right)^{{3}/{5}}
\left( \frac{f_{\rm mag}}{0.1}\right)
\left( \frac{f_{\rm rel}}{0.1}\right)
\left( \frac{\nu}{1{\rm GHz}}\right)^{-3}
.
\end{equation}

%

In this paper, we take the assumption about the directions of magnetic
fields,  $\sin(\theta_{\rm
B})\sim 1$, in all SN remnants for simplicity.
This implies that we have ignored the line of sight magnetic field
and consequently the FR effects associated to it. 
This assumption is consistent in our case since the FR effects in a
relativistic plasma decreases as the plasma becomes more 
relativistic~\citep{beckert}. However, in a $non$-$relativistic$ plasma,
the FR effect is not negligible in order to evaluate the resultant
circular polarization via the FC and it is necessary to solve
a full set of the polarization transfer equation including the FR and FC simultaneously. 

\subsection{Power spectrum of the FC induced by Pop~III stars}

In the $\Lambda$CDM model, Pop III stars are predicted to have formed
inside dark matter halos. Therefore, in order to calculate $P_\alpha$,
we can evaluate $P_\alpha$, based on the halo model~\cite{Seljak:2000gq}.
For simplicity, we assume that Pop III stars form in halos with the
virial temperature $T_{\rm vir} > 10^4~$K where atomic hydrogen cooling is
effective for the collapse. We also assume that {\it one} Pop III star is formed per halo
and consequently, one halo hosts one Pop III SN remnant.
Accordingly, we can write the power spectrum for the FC from Pop III
star SN remnants in the following way, 
\begin{equation}
\label{eq:palpha}
P_{\alpha}(z,k)
=P_{\rm mat}(z,k)\left |
\int_{M_{\rm thr}} dM~ \frac{dn}{dM} b(M,z) \tilde{\alpha}_0(z,k)
\right |^2,
\end{equation}
where 
$P_{\rm mat}(z,k)$ is the linear matter power
spectrum at a redshift $z$, $b(M, z)$ is the linear bias of a dark matter halos,
$dn/dM$ is the
mass function,
and
$\tilde{\alpha}_0$ is the Fourier component of $\alpha_0$
and given by, 
\begin{equation}
 \tilde{\alpha}_0(z,k)= \int \frac{d^3\vec{r}}{(2 \pi)^3} \alpha_0(z) p_h(
\vec{r})
  \exp({-i\vec{k}\cdot\vec{r}}).
\end{equation}
where $p_h(\vec{r})$ is the profile of one SN remnant.
For simplicity, we assume that the FC rate is homogeneous inside the SN
remnant and the profile is given by $p_h(\vec{r}) =
1$ for $r_p <|\vec{r}| <r_s$, otherwise $p_h =0$.

In Eq.~(\ref{eq:palpha}),
we consider only the halo-halo 
correlation term of the halo model, and we neglect the one-halo
Poisson term. In other words, we take into the correlation between SN remnants
in different halos, while we ignore the correlation within a given SN remnant. 
In this work, we are interested in the correlation of the CP on large
scales~(the order of 10~Mpc or equivalently $l<2000$ in terms of multipole).
Since the typical size of SN remnants is much smaller, we can neglect
the correlation contribution within a given SN remnant.
We will discuss in more detail later.

In Eq.~(\ref{eq:palpha}),~${M_{\rm thr}}$ is the threshold mass of the
halos hosting a Pop III star and 
corresponds to the virial mass with $T_{\rm vir} =10^4~$K~\cite{Barkana:2000fd},
\begin{equation}
\label{eq:virial}
M_{\rm thr} = 3.5 \times 10^7 h^{-1}  \left(\frac{T_{\rm vir}}{10^4 ~{\rm K}} \right)^{3/2}
\left(\frac{\Omega_m}{\Omega_m(z)}\frac{\Delta_c}{18\pi^2}\right)^{-2}
\left( \frac{1+z}{10} \right)^{-3/2} ,
\end{equation}
where
$\mu$ is the mean molecular weight and $m_{p}$
is the proton mass. For a fully ionized plasma, we used, $\mu=0.6$.

We choose the mass function, $dn/dM$, to be alternately defined in terms
of a multiplicity function $f(\nu)$ as,
\begin{equation}
\label{eq:fnu}
\frac{dn}{dM}dM=\frac{\bar{\rho}}{M}f(\nu)d\nu.
\end{equation}
Here $\bar{\rho}$ is the mean matter density in the universe
and $\nu$ is defined as $\nu=\left(\delta_c/\sigma_M \right)^2$
where $\delta_c$ is the threshold overdensity of spherical collapses at
redshift $z$ and
$\sigma_{M}$ is the rms linear density fluctuations obtained with a
top-hat filter of mass $M$ at an initial time (see \cite{Seljak:2000gq} and references therein).

For the function $f(\nu)$, we adopt the function proposed by Sheth and
Tormen~\cite{Sheth:1999mn}
\begin{equation}
\nu f(\nu) =
 A(1+\nu^{-p}_1)\left(\frac{\nu_1}{2 \pi}\right) ^{1/2}
 e^{-\nu_1/2},
 \label{eq:sheth-tormen}
\end{equation}
where $\nu_1=a\nu$ with $a=0.7$, $p=0.3$
and $A$ is the normalization constant determined by $\int f(\nu)d\nu=1$.
Following Ref.~\cite{Seljak:2000gq}, the linear bias, $b_{\nu}$, 
is given in terms of $\nu$ by
\begin{equation}
\label{eq:bnu}
b(\nu)=1+\frac{\nu-1}{\delta_c}\frac{2p}{\delta_c(1+\nu_1^p)}.
\end{equation}

\begin{figure}[tbp]
\includegraphics[height=0.5\textwidth]{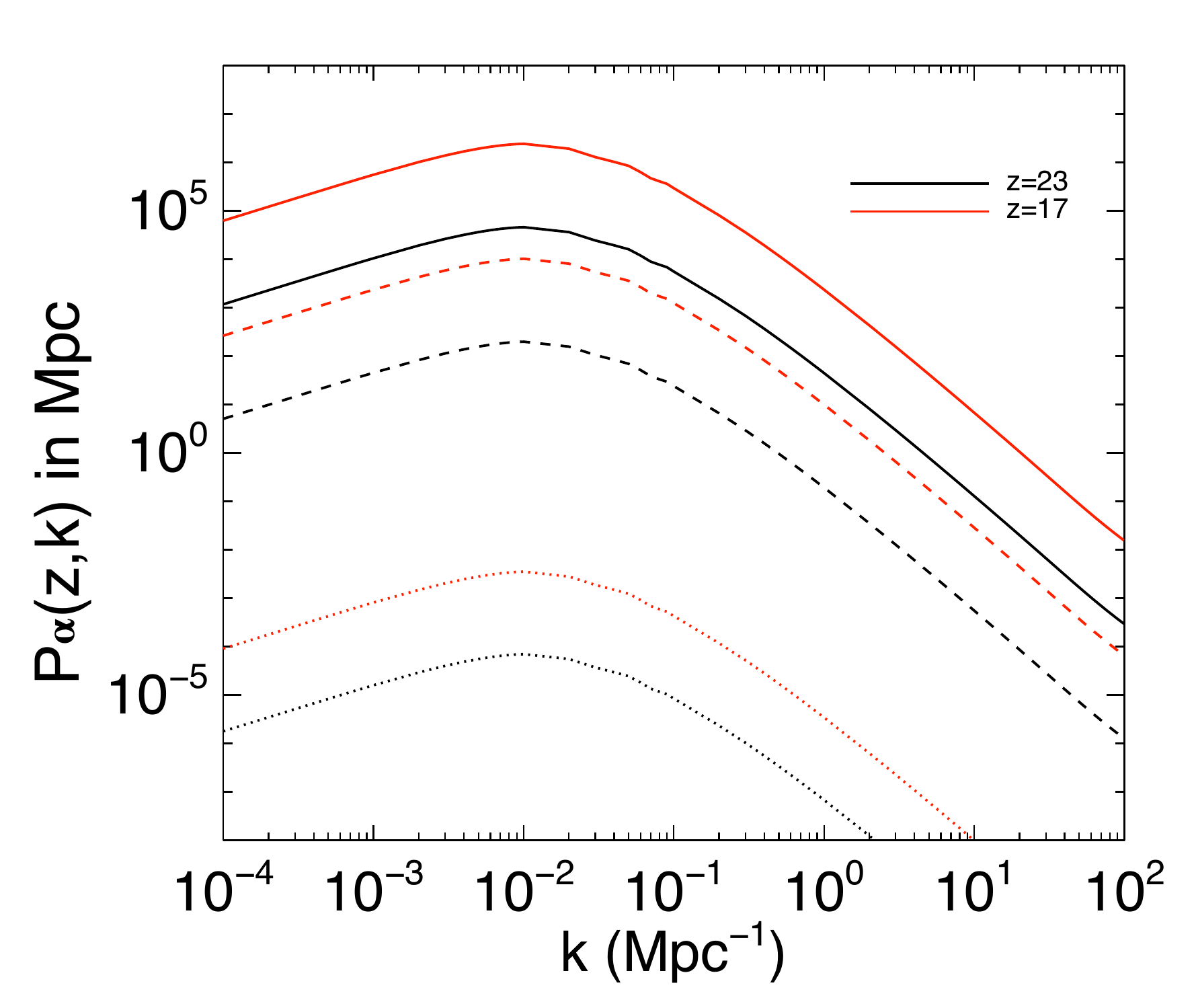}
\caption{Power spectra of Faraday conversion generated by the First
stars at different redshifts. The unit of $P_{\alpha}(k,z)$ is
in Mpc.
 The solid lines correspond to $t_{\rm age}=10^4$ years and
$\nu=1$ GHz. The dashed lines correspond to $t_{\rm age}=10^5$ years and
$\nu=1$ GHz. Lastly, the dotted lines correspond to $t_{\rm age}=10^4$
years and $\nu=30$ GHz. 
Here, $\nu$ is the frequency of the CMB photons observed today.}
\label{fig:palpha}
\end{figure}

In Fig.~\ref{fig:palpha}, we plot the power spectrum of $\alpha$, $P_{\alpha}(z,k)$.
As the redshift decreases, the power spectrum amplitude grows. This 
is simply due to the growth of the density fluctuations, as a result of
which the resultant number density
of halos becomes larger at lower redshifts.
Fig.~\ref{fig:palpha} also shows that the power spectrum depends on
the age of SN remnants $\tage$ and the observation frequency $\nu$.

An estimation of the power spectrum for $\alpha$ in Fig.~\ref{fig:palpha}
at the peak scale ($k$ $\sim$ 10$^{-2}$Mpc$^{-1}$) can be verified as
follows. For t$_{\rm age}$=10$^5$ yr and $\nu$=1 GHz, we obtain the
value of $\tilde{\alpha}_0 \sim$ 0.2 Mpc$^2$. At $z \sim 20$, 
the integral of $dn/dM$ over M in
Eq.~(\ref{eq:palpha}) gives $\sim$ 1 Mpc$^{-3}$~\citep{Barkana:2000fd},
and the bias factor is $b(M,z) \sim$ 1.
Since the amplitude of the matter power spectrum at $k \sim 10^{-2}$
is $P(k) \sim 100~{\rm Mpc}^{3}$,
we obtain $P_\alpha \sim 10~$Mpc at the peak scale from 
Eq.~(\ref{eq:palpha}).
%

Since we consider only the halo correlation term, as
shown in Eq.~(\ref{eq:palpha}),
the spectral shape is same as the one of the linear matter power spectrum.
However, in general, the Poisson term is non-negligible on small
scales. Taking into account the Poisson term, one can expect
that the shape of the power spectrum $P_\alpha$ would deviate from the linear
matter power spectrum and be enhanced around $k >10^5$ Mpc$^{-1}$, which also corresponds to the
typical scale of the SN remnants of Pop III stars.

Since we consider only the halo correlation term, as
shown in Eq.~(\ref{eq:palpha}),
the spectral shape is same as the one of the linear matter power spectrum.
However, in general, the Poisson term is expected to be non-negligible on small
scales.
Taking into account the Poisson term, one can expect
that the shape of the power spectrum $P_\alpha$ would deviate from the linear
matter power spectrum and be enhanced around the scale corresponding to the
typical scale of the SN remnants of Pop III stars.
The average proper size of the remnants
of age, $t_{\rm age}$=10$^6$ years is $\sim 10$~pc at $z \sim 20$. This
corresponds to an apparent angular size of $\sim 10^{-6}$ in degrees.
Therefore, the Poisson term could contribute much on angular power
spectrum around multipoles of $l \sim 10^6$.
Since such scales are too small to observe the angular power spectrum,
we do not address the Poisson term in this paper.

%

\section{Results for the predicted $\clvv$ from Pop III stars}
\label{sec:results}

We numerically calculate the angular power spectrum of the CMB CP due to
the SN remnants of Pop III stars,
substituting the power spectrum $P_\alpha$ obtained in the previous section to Eq.~(\ref{eq:clvv}).

Although Eq.~(\ref{eq:clvv}) involves multiple summations of multipoles,
we can reduce the calculation.
Due to the property of the Wigner-$3j$ symbols in Eq.~(\ref{eq:complicated}), the non-zero
contributions come from the terms with $l'+L$ is $odd$ in
Eq.~(\ref{eq:clvv}). Therefore, non-vanishing $I_{lm}$ requires $m'\neq
0$ in Eq.~(\ref{eq:clvv}).
Under the assumption that the CMB is statistically isotropic, the
angular correlation of multipole components $V_{lm}$ defined in
Eq.~(\ref{eq:dec_V}) is independent of $m$. Therefore the calculation with only $m=0 $ in Eq.~(\ref{eq:clvv}) is enough to obtain
the angular power spectrum.
Additionally, in order to reduce computational efforts,
we ignore the azimuthal dependence due to $m''$. In other words we just 
multiply $(2l''+1)$ instead of evaluating a $m''$ dependent summation.

\begin{figure}[tbp]
\includegraphics[height=0.5\textwidth]{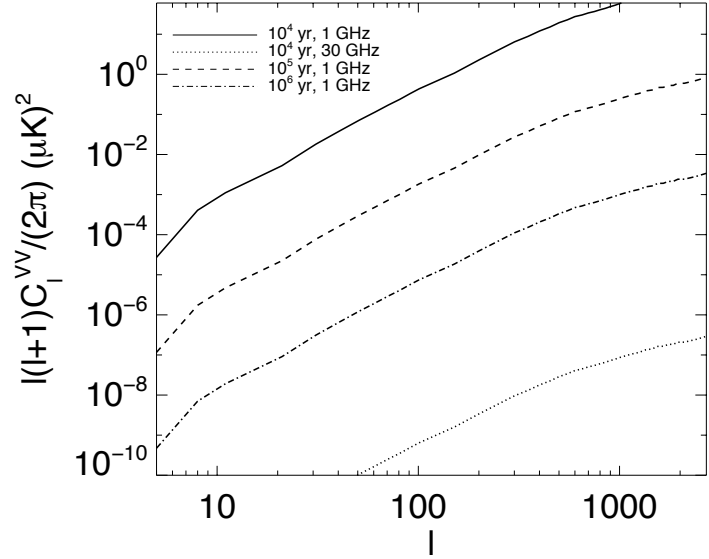}
\caption{Angular power spectra of circular polarization from
Eq.~(\ref{eq:clvv}) times $T_{\rm CMB}^2$ in ($\mu$K)$^2$. We have chosen
a few cases for the age of the remnant of the Pop III explosion, with
$10^4<t_{\rm age}<10^6$ in years. The frequency of observation of the
CMB has been chosen to be $\nu=$ 1 and 30 GHz.  Other parameters for
this plot are described in Sec.~\ref{sec:results}. In this figure only
$one$ Pop III star was assumed per halo.}
\label{fig:clvv}
\end{figure}

In Fig.~\ref{fig:clvv}, we present the results of $T_{CMB}^2\clvv$ in the units of $(\mu
K)^2$. Here we assume that Pop III stars exist in high redshifts between
$z=24$ and $17$, and we set
$\nu=1$~GHz, $f_{\rm rel}=0.1$ and $f_{\rm mag}=0.1$. Although
$P_\alpha$ has a peak at $k \sim 10^{-2}$ corresponding to $l \sim 100$,
the angular power spectrum of the CMB CP $\clvv$ peaks at higher $l \sim
2000$. This is because $\clvv$ is the convolution between the $P_\alpha$
and the E-mode power spectrum. We also show the dependence on $\tage$ in Fig.~\ref{fig:clvv}.As
the SN remnants expand with $\tage$ increase, the number density of
relativistic electrons and magnetic field energy decrease. Accordingly
the FC becomes ineffective in the case with large $\tage$.
For comparison, we also plot the angular power spectrum with $\nu =
30~$GHz. The spectrum is strongly sensitive to the frequency of the CMB
observation. Although the frequency dependence depends on the power-law
index $\gamma$ of the relativistic electron distributions, the amplitude of
$\clvv$ is proportional to $\nu^{-6}$ for the case with $\gamma =2$.

One can roughly estimate $\clvv$ in the following way.
$\clvv$ is a
convolution between power spectrum of circular polarization coefficient
$\alpha$ and that of E-mode polarization and can be approximated to
$l^2 \clvv \sim l^2 C_{l}^{\alpha \alpha} C_l^{EE}$ where
$C_{l}^{EE}$ and
$C_l^{\alpha \alpha}$ are the angular power spectra for CMB E-mode
polarization and the FC parameter $\alpha$, respectively.
We are interested in scales $l>100$. 
The Limber approximation gives the
relation between $C_l^{\alpha \alpha}$ and $P_\alpha$ (given in
Fig.~\ref{fig:palpha}) as
$l^2 C_l^{\alpha \alpha} \sim k P_\alpha (l/r_z)$ where $r_z$ is the
radial distance to redshift $z$ and $kr_z \sim l$. For example, at $z$ $\sim$ 20, 
in scales $l\sim 10^3$, $l/r_{z \sim 20}$ corresponds to~$k \sim 0.1~{\rm Mpc}^{-1}$.
For $\tage$=10$^5$ yr and $\nu$=1 GHz, including the contributions from
all redshifts, at $k \sim 0.1$ Mpc$^{-1}$, $P_\alpha \sim 10^3~{\rm Mpc}$ from
Fig.~\ref{fig:palpha}.
The angular power spectrum of E-mode polarization is
$C_l^{EE} \sim 10^{-2}~{\rm \mu K}^2$ at $l\sim 10^3$.
Therefore, the resultant angular power spectrum of the CP is
$l^2 C_l^{VV} \sim k P_\alpha C_l^{EE} \approx 10^{-2}~{\rm \mu K}^2$ for 
$\tage=10^5~\rm yr$ at 1~GHz.
This estimation is
consistent with our calculated total $\clvv$ in Fig.~\ref{fig:clvv}.
%

\begin{figure}[tbp]
\includegraphics[height=0.5\textwidth]{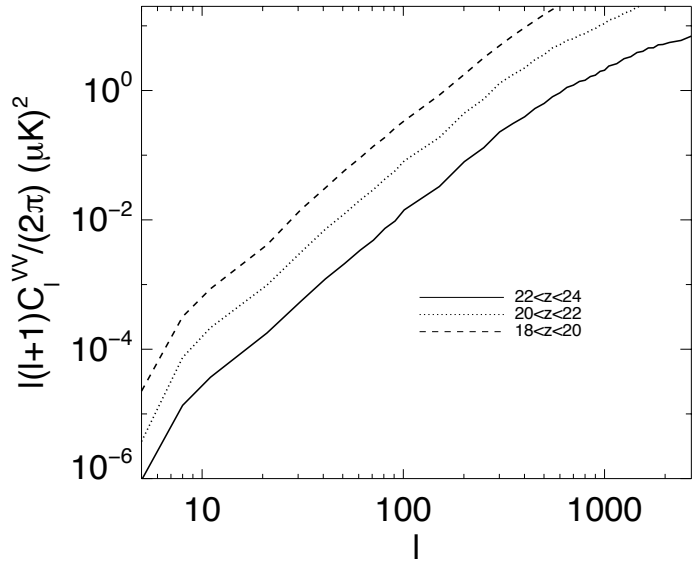}
\caption{Angular power spectra of circular polarization from
Eq.~(\ref{eq:clvv}) times $T_{\rm CMB}^2$ in ($\mu$K)$^2$. We have chosen
a few redshift slices and the age of the remnant of the Pop III
explosion to be $t_{\rm age}=10^4$ in years. 
The frequency of observation of the
CMB has been chosen to be $\nu=$ 1 GHz for this plot. Other parameters for
this plot are described in Sec.~\ref{sec:results}.}
\label{fig:clvv-slices}
\end{figure}

In Fig.~\ref{fig:clvv-slices}, we show the redshift dependence of
$\clvv$. In this case we have only considered remnants with t$_{\rm
age}$=10$^4$ yrs. Fig.~\ref{fig:clvv-slices} conveys that $\clvv$ decreases with
increasing redshift. This is primarily due to the increase in the number
density of Pop III stars in lower redshifts directly due to increase in
the number of halos.
In this paper, although we do not take into account the redshift dependence of
the evolution of 
Pop III star properties, their properties strongly depend on
redshift. In particular, the metal pollution due to explosions of Pop III
stars make mass of Pop III stars small and, finally, the abundance of Pop III stars are
dominated by Pop II stars. These redshift evolutions induce the
suppression in the efficiency of the FC in lower redshifts. Therefore,
$\clvv$ in lower redshifts is expected to lower than our
estimation. However, this is beyond the scope of this paper.

There are two factors involved in the understanding an order of magnitude
estimate for the redshift dependence. One, an increase in the number
density of halos with decreasing redshift and two, a given length scale  
is manifested at a slightly larger angular scale (or at a
slightly higher multipole) with decreasing redshift. The dominating
effect of increase in the signal due to increase in the number density
of halos with decreasing redshift is clearly manifested in
Fig.~\ref{fig:clvv-slices}. An 
approximate increase of two orders of magnitude in the power spectrum at a given
scale upon decrease of redshift from $z=23$ to $z=17$ is shown in
Fig.~\ref{fig:clvv-slices}. This change is also
manifested in Fig.~\ref{fig:clvv-slices} with a similar reduction in the
signal between $z=23$ and $z=17$. 

\section{Discussion and Summary}

In this paper, we have studied the CP signals of the CMB, 
focusing on the Faraday Conversion~(FC) caused by the SN remnant of the Pop III star explosions.
In the SN remnant, the relativistic electrons are produced and the
magnetic fields are amplified. Therefore, during the propagation through
the SN remnant, the CMB undergoes FC which transfers some of its linear
polarization into circular polarization (CP).

In this paper, 
we have derived an analytic form for the angular power spectrum of
the CMB CP. This analytic form is general to the CP due to the FC. The
angular power spectrum is the convolution between the CMB E-mode
and the FC rate power spectra.

We have applied this analytic form to estimate
the CP of the CMB due to the SN remnants of Pop III stars.
Compared with primary E-mode polarization, the amplitude of the produced
CP is suppressed by a factor of 10$^{-4}$ in terms of the angular power
spectrum for remnants of the Pop III stars with
t$_{\rm
age} >10^5$ yr. The efficiency of the FC strongly depends on the frequency of CMB photons.
We found that the amplitude of the CP angular power spectrum is
proportional to $\nu^{-6}$. 
The signals of the CP fall off with increasing frequency.
%
As the SN remnant evolution, we adopt the simple analytic model, the
Sedov similar solution.
In this model,
as the SN remnant evolves, the number density of relativistic
electrons and the amplification of magnetic fields are suppressed.
Therefore, the signals of the CP also decrease with the age of the
remnant, $\tage$ growing.

Throughout this paper,
we have assumed that energy of each Pop III explosion is $E_{\rm SN}=10^{53}$
ergs. However, $E_{\rm SN}$ depends on various properties of Pop III stars and 
is theoretically uncertain.
However, the produced CP also depends on the energy of the explosion.
The angular power spectrum $\clvv$ is roughly proportional to $E_{\rm
SN}^{(16+2\gamma)/10}$ with the spectral index $\gamma $ of the
relativistic electron distribution. For $\gamma =2$,
we retrieve
an approximate quadratic dependence of $\clvv$ on $E_{\rm SN}$. 
Therefore, the detailed energy distribution of Pop III SNe is required to predict $C_l^{VV}$
precisely.
Additionally, there exists the theoretical uncertainty on the
conversion of the SN energy into the energies of relativistic electrons
and magnetic fields, which we parameterize as $f_{\rm rel}$
and $f_{\rm mag}$.
We will address the modeling of these parameters
based on simulations in a future work.

%

We have assumed the
direction of the magnetic field generated by their explosions, 
to be aligned to the $\hat{z}$ axis. In
other words, we have assumed $B_{\rm mag}=B_z$. If we relax this assumption,
and consider a general direction of $B_{\rm mag}$, we may use $B_z=B_{\rm mag}/{\sqrt{3}}$ according to the
equipartition over the all directions.
The signal of CP due to only $B_z$ is then suppressed by a factor of 1/9. In this case, there are additional
contribution due $B_x$ and $B_y$, the components of magnetic field in
the $\hat{x}$ and $\hat{y}$ directions. These contributions could be
additive or subtractive and a precise estimate is difficult 
without a more detailed numerical modeling.
%
%



In our evaluation of the CMB CP, we ignore the Faraday 
rotation~(FR) effect.
The FR arises when the CMB photons pass through a
magnetised plasma.
Therefore, when the FC of the CMB occurs, the FR is also
expected to be effective.
Since the FR rotates the direction of the linear polarization,
the relative angle between the linear polarization and the magnetic
fields changes. Even more, when the FR is efficient,
it might be change the sign of the CP due to the FC.
Therefore the magnitude and sign of the final CP might depend on the details
of the FR in the system.
Although the FR due to magnetic fields in intergalactic medium, galaxy
clusters and the Milky way has been
studied~(e.g.~\cite{Kosowsky:1996yc,Tashiro:2007mf,De:2013dra}), the FR of the CMB in SN
remnants of Pop III stars has not been addressed yet. In order to evaluate the effect of
SN remnants on the CMB
linear and circular polarization, it is required to study the FR and FC
consistently. 
We propose to investigate this detail in the future.

%

In Ref.~\cite{Cooray:2002nm}, the signals of the CMB CP due to the propagation
through galaxy clusters are predicted that the peak amplitude is
$10^{-1}$~($\mu$K)$^2$ at $l=10^3$ for $\nu = 1$GHz.
Our predicted signals due to the Pop III
stars with $t_{\rm age } =10^4$~yr can dominate these signals. 
Although the signals due to the Pop III
stars with $t_{\rm age } =10^5$~yr are comparable with these signals around
$l \lesssim 10^3$, these can dominate the signals from galaxy
clusters on the smaller scales. Even with $t_{\rm age} =10^6$~yr, the
signals from the Pop III SNe would dominate the ones from galaxy
clusters at $l>5 \times 10^3$.
Therefore, the Pop III SNe can contribute the CMB CP
significantly, in particular, on small scales $l>10^3$.

%
%

We propose that if the future CMB experiments are equipped with CP
measuring instruments, the CMB observation can also be used as a probe
of the Pop III stars. In addition to the $\langle VV\rangle$
correlations, $\langle TV \rangle$ and
$\langle EV \rangle$ correlations are expected to be higher, leading up to a better
detection prospect. An unique frequency signature of the CP signal due to the 
FC and absence of any significant foreground makes CP
signal to be a promising probe of the Pop III stars, which are yet 
unobserved.

\acknowledgements 
We thank Eiichiro Komatsu for helpful comments and Tanmay Vachaspati 
for important discussions. We thank 
Frank Timmes for helping us with the computing
resources for our calculations.
We are also grateful for computing resources at the ASU Advanced
Computing Center (A2C2). We convey sepcial thanks to Philip Lubin and
Torsten Ensslin for their helpful comments on the understanding of the
physics of the Faraday conversion mechanism.
SD is supported by a NASA Astrophysics
theory grant NNX11AD31G and HT is supported
by DOE at the Arizona State University.



\newpage
\appendix
\section{Derivation of $\clvv$}
\label{app:b}

In this appendix, we derive the angular power spectrum of the CP of the
CMB, Eq.~(\ref{eq:clvv}). As shown in Eq.~(\ref{eq:eq_V}), the observed
Stokes parameter $V$ due to the FC is given by
\begin{equation}
\label{eq:eq_Vapp}
V(\hat{n})=-2\int_{r_*}^{0}dr ~ U(r, \vec{x},\hat{n})
\alpha(r,\vec{x},\hat{n}, \hat{b}).
\end{equation}

According to Eq.~(\ref{eq:alpha}), $\alpha $ is proportional to $(\sin
\theta_B )^{(\gamma+2)/2}$ where $\theta_B$ is the angle between the
line-of-sight direction $\hat{n}$ and the direction of magnetic fields
$\hat b$.
In order to express the $\theta_B$-dependence explicitly, we rewrite $\alpha$ as
\begin{equation}
\label{eq:spin0-spin2}
\alpha(r,\vec{x},\hat{n}, \hat{b}) =
\alpha_0(r,\vec{x})
(\sin \theta_B)^{(\gamma+2)/2} .
\end{equation}
For simplicity, 
we consider only $z$-direction component of
magnetic fields with adopting $\gamma =2$.
In this case, the $\theta_B$-dependence can be written as
\begin{equation}
(\sin \theta_B)^{(\gamma+2)/2}  =
 \sin ^2\theta = \frac{1}{2} \sqrt{\frac{32 \pi}{15}} 
\left(
 {}_{2}Y_{2}^0( \hat n)+ {}_{-2}Y_{2}^0(\hat n)
\right),
\end{equation}
where $\theta$ is a polar angle component of $\hat n$ in a spherical coordinate system
$\hat n = (\theta, \phi)$ for the sky.
We perform the Fourier decomposition of $\alpha_0(r,\vec{x})$,
\begin{equation}
\label{eq:alpha-fourier}
 \alpha_0(r,\vec{x}) = \int \frac{d^3k}{(2 \pi)^3} \tilde \alpha (r, k)
  \exp({i\vec{k}\cdot\vec{x}}).
\end{equation}
Accordingly, $\alpha(r,\vec{x},\hat{n}, \hat{b})$ is expressed as
\begin{equation}
\label{eq:fcrate}
\alpha(z,\vec{x},\hat{n}, \hat{b}) 
=
 2\pi  \sqrt{\frac{32 \pi}{15}} 
\int
\frac{d^3k}{(2\pi)^3} {\tilde \alpha}(z,k)
\left(
 {}_{2}Y_{2}^0( \hat n)+ {}_{-2}Y_{2}^0(\hat n)
\right)
\sum_l
(-i)^lj_l(kr)\sum_{m=-l}^{m=l}Y^*_{lm}(\hat{k})Y_{lm}(\hat{n}),
\end{equation} 
where we use the Rayleigh expansion,
\begin{equation}
\label{eq:planewave}
e^{i\vec{k}.\vec{x}}=\sum_{lm}4 \pi (-i)^l j_{l}(kr) Y_{l}^m (\hat{k})
Y_{l}^m ( \hat{n}),
\end{equation}
with $\hat k = \vec{k}/k$.

From Eq.~(\ref{eq:u_dec}),
we can write the Stokes parameter $U$ as
\begin{equation}
U(r, \vec{x},\hat{n}) =-\frac{i}{2}\int \frac{d^3k}{(2\pi)^3} \sum_l
\sum_{m=-2}^{m=2} E_{lm}\left( _{2}G_l^m -_{-2}G_{l}^m \right).\label{eq:U_fr}
\end{equation}

Substituting Eqs.~(\ref{eq:fcrate}) and (\ref{eq:U_fr}) to Eq.~(\ref{eq:eq_Vapp}) provides
\begin{eqnarray}
V(\hat{n})
 &=&
- 8\pi^2 i\sqrt{\frac{32 \pi}{15}}
\sum_l \sum_{m=-2}^{m=2}
\sum_{l'}\sum_{m'=-l'}^{m'=l'}
\sum_{l''}\sum_{m''=-l''}^{m''=l''}
\int_{r_*}^{0} dr
~\int \frac{d^3k}{(2\pi)^3}
\int \frac{d^3k'}{(2\pi)^3}
\nonumber \\
&&\times
(-i)^{l+l'+l''} \sqrt{\frac{4\pi}{2l+1}} {\tilde \alpha}(r,k')
E_{lm}(k,r)
j_{l'}(k'r) j_{l''}(kr)
Y^*_{l'm'}(\hat{k}')
Y^*_{l''m''}(\hat{k})
\nonumber \\
&&\times
\bigl( _{2}Y_l^m (\hat n)-_{-2}Y_{l}^m (\hat n) \bigr)
\left(
 {}_{2}Y_{2}^0( \hat n)+ {}_{-2}Y_{2}^0(\hat n)
\right)
  Y_{l'm'}(\hat{n})
  Y_{l''m''}(\hat{n}).
\end{eqnarray}

Decomposing the Stokes parameter $V$ to spherical harmonics as shown
Eq.~(\ref{eq:dec_V}), we obtain
\begin{eqnarray}
 V_{lm} &= &
- 8\pi^2 i\sqrt{\frac{32 \pi}{15}}
\sum_{l'} \sum_{m'=-2}^{m'=2}
\sum_{l''}\sum_{m''=-l''}^{m''=l''}
\sum_{l'''}\sum_{m'''=-l'''}^{m'''=l'''}
\int_{r_*}^{0} dr
~\int \frac{d^3k}{(2\pi)^3}
\int \frac{d^3k'}{(2\pi)^3}
\nonumber \\
&&\times
(-i)^{l'+l''+l'''} \sqrt{\frac{4\pi}{2l'+1}} {\tilde \alpha}(r,k')
E_{l'm'}(k)
j_{l''}(k'r) j_{l'''}(kr)
Y^*_{l''m''}(\hat{k}')
Y^*_{l'''m'''}(\hat{k})
\nonumber \\
 &&\times
  \int d^2 {\hat n}~
\bigl( _{2}Y_{l'}^{m'} (\hat n)-_{-2}Y_{l'}^{m'} (\hat n) \bigr)
\left(
 {}_{2}Y_{2}^0( \hat n)+ {}_{-2}Y_{2}^0(\hat n)
\right)
  Y_{l''m''}(\hat{n})
  Y_{l'''m'''}(\hat{n})
    Y_{lm}(\hat{n}).
\label{eq:V_lm}
\end{eqnarray}

Since the product of two spherical harmonics can be
given in terms of 3$j$ symbols, the non-zero contributions comes from
the terms including  $  _{2}Y_{l}^{m} (\hat n) {}_{-2}Y_{2}^0( \hat n)$
and $_{-2}Y_{l}^{m} (\hat n) {}_{2}Y_{2}^0( \hat n)$ which are
represented as
\begin{eqnarray}
  _{2}Y_{l}^{m} (\hat n) {}_{-2}Y_{2}^0( \hat n)
&=& \sum_{L}
 \sqrt{\frac{5(2 l +1) (2 L+1)} {4\pi}}
\left(
\begin{array}{ccc}
l & 2 & L \\
m & 0 &m  \\
\end{array}
\right)
\left(
\begin{array}{ccc}
l & 2 & L \\
2 & -2 & 0  \\
\end{array}
\right) Y_L ^0 (\hat n),
\nonumber \\
 _{-2}Y_{l}^{m} (\hat n) {}_{2}Y_{2}^0( \hat n)
&=& \sum_{L}
 \sqrt{\frac{5(2 l +1) (2 L+1)} {4\pi}}
\left(
\begin{array}{ccc}
l & 2 & L \\
m & 0 &m  \\
\end{array}
\right)
\left(
\begin{array}{ccc}
l & 2 & L \\
-2 & 2 & 0  \\
\end{array}
\right) Y_L ^0 (\hat n).
\end{eqnarray}

Integrating Eq.~(\ref{eq:V_lm}) over $\hat n$ yields
\begin{eqnarray}
 \label{eq:v_lm_2}
 V_{lm} &= &
- 8\pi^2 i\sqrt{\frac{32 \pi}{15}}
\sum_{l'} \sum_{m'=-2}^{m'=2}
\sum_{l''}
\sum_{l'''}\sum_{m''+m'''=m}
\int_{r_*}^{0} dr
~\int \frac{d^3k}{(2\pi)^3}
\int \frac{d^3k'}{(2\pi)^3}
\nonumber \\
&&\times
(-i)^{l'+l''+l'''} \sqrt{\frac{4\pi}{2l'+1}} {\tilde \alpha}(r,k')
E_{l'm'}(r,k)
j_{l''}(k'r) j_{l'''}(kr)
Y^*_{l''m''}(\hat{k}')
Y^*_{l'''m'''}(\hat{k})
I_{lm}.
\end{eqnarray}
Here $I_{lm}$ represents the integration over $\hat n$,
\begin{flalign}
I_{lm}=  & \int d^2 {\hat n}~
\left( _{2}Y_{l'}^{m'} (\hat n) {}_{-2}Y_{2}^0(\hat n)-_{-2}Y_{l'}^{m'} (\hat n) 
 {}_{2}Y_{2}^0( \hat n)
\right)
  Y_{l''m''}(\hat{n})
  Y_{l'''m'''}(\hat{n})
    Y_{lm}(\hat{n})
\nonumber \\
 & \quad=\sum_{L, L'}
 \frac{(2L+1) (2 L' +1)}{4\pi}
 \sqrt{\frac{5(2l+1) (2 l' +1) (2l''+1) (2l'''+1) } {4\pi}}
\nonumber \\
 & \qquad \times
\left[
\left(
\begin{array}{ccc}
l' & 2 & L \\
2 & -2 & 0  \\
\end{array}
\right)
 -
 \left(
\begin{array}{ccc}
l' & 2 & L \\
-2 & 2 & 0  \\
\end{array}
\right)
\right]
 \left(
\begin{array}{ccc}
l' & 2 & L \\
m' & 0 &-m'  \\
\end{array}
\right)
\nonumber \\
 & \qquad \times
  \left(
\begin{array}{ccc}
l'' & l''' & L' \\
m'' & -m'' & 0  \\
\end{array}
\right)
   \left(
\begin{array}{ccc}
L & L' & l \\
-m & 0 & m  \\
\end{array}
\right)  \left(
\begin{array}{ccc}
l'' & l''' & L' \\
0& 0 & 0  \\
\end{array}
\right)
   \left(
\begin{array}{ccc}
L & L' & l \\
0 & 0 & 0  \\
\end{array}
\right),
\label{eq:complicated}
\end{flalign}
where, for non-zero $I_{lm}$, $m'' + m'''$ should be $m$.
Note that in the above expression, $m'=0$ is not allowed since it
requires $l'+L=$even, which is opposed by the first difference term in
Eq.~(\ref{eq:complicated}) that requires $l'+L$ to be odd in order to
have a non-vanishing difference term.

Let us calculate the angular power spectrum of $V_{lm}$, plugging
Eq.~(\ref{eq:v_lm_2}) to Eq.~(\ref{eq:def_clvv}),
\begin{eqnarray}
\clvv
&=&
64 \frac{32 \pi^5}{15}
\sum_{l' l'' l'''} \sum_{m'=-2}^{m'=2} \sum_{m''+m'''=m}
\sum_{L'L''L'''} \sum_{M'=-2}^{M'=2}
\sum_{M''+M'''=m}
\int dr \int dr'
~\int \frac{d^3k}{(2\pi)^3}
\int \frac{d^3k'}{(2\pi)^3}
\int \frac{d^3K}{(2\pi)^3}
\int \frac{d^3K'}{(2\pi)^3}
\nonumber \\
&&\times
\sqrt{\frac{(4\pi)^2}{(2l'+1)(2L'+1)}}
\langle 
{\tilde \alpha}(r, k') {\tilde \alpha}(r',K')
\rangle
\langle
E_{l'm'}(k,r)
E_{L'M'}(K,r')
\rangle
\nonumber \\
&& \times
j_{l''}(k'r) j_{l'''}(kr)
j_{L''}(K'r') j_{L'''}(Kr')
Y^*_{l''m''}(\hat{k}')
Y^*_{l'''m'''}(\hat{k})
Y^*_{L''L''}(\hat{K}')
Y^*_{L'''L'''}(\hat{K})
I_{lm}(l')
I_{lm}(L')
\nonumber \\
 &=&
64 \frac{32 \pi^5}{15}
\sum_{l' l'' l'''} \sum_{m'=-2}^{m'=2} \sum_{m''+m'''=m}
\int dr \int dr'
~\int \frac{k^2 dk}{(2\pi)^3}
\int \frac{k'^2 dk'}{(2\pi)^3}
\nonumber \\
&&\times
\frac{4\pi}{(2l'+1)}
\langle 
{\tilde \alpha}(r, k') {\tilde \alpha}(r',K')
\rangle
\langle
E_{l'm'}(k,r)
E_{L'M'}(K,r')
\rangle
j_{l''}(k'r) j_{l'''}(kr)
j_{l''}(k'r') j_{l'''}(kr')
I_{lm}^2.
\label{124509_24Dec13}
\end{eqnarray}

Since we are interested in the scales corresponding to $l>100$,
we can apply the Limber approximation to Eq.~(\ref{124509_24Dec13}),
\begin{equation}
 \int k^2 dk P(k) j_{l} (kr) j_{l} (kr') \approx \frac{ \pi
  \delta(r-r')}{2 r^2} P(k) |_{k=l/r}.
\end{equation}
Finally, we obtain the angular power spectrum of the Stokes
parameter $V$ as
\begin{eqnarray}
\clvv=
 \frac{128}{15}
\sum_{l' l'' l'''} \sum_{m'=-2}^{m'=2} \sum_{m''+m'''=m}
\int dr 
\int 
\frac{k^2 dk}{(2l'+1) r^2}
P_\alpha
\left(\frac{l''}{r},r \right)
P_{E_{lm}}
\left( k,r \right)
j_{l'''}^2(kr) 
I_{lm}^2,
\end{eqnarray}
where
we define the power spectra of $\alpha $ and $E_{lm}$ at a comoving
distance $r$
as
\begin{equation}
 \langle \alpha ({\bm k}, r)
  \alpha^*({\bm k}', r) \rangle= (2 \pi)^3
  \delta^3(\bm{k}-\bm {k}')
  P_\alpha(k,r), \quad
  \langle E_{lm}({\bm k},r) E_{lm}^*({\bm k}',r) \rangle= (2 \pi)^3
  \delta^3({\bm k}-{\bm k}') P_E(k,r).
\end{equation}

In our final calculation of
$\clvv$, we have simplified the evaluation of $I_{lm}$  by using
Eq.~(B2) and Eq.~(B4) of Ref.~\cite{tina} for Wigner-3$j$s with zero azimuthal
numbers. For the Wigner-3$j$s with non-zero azimuthal numbers, we have
used Eq.~(A5c) of Ref.~\cite{sprung} wherever appropriate.

\end{document}